\documentclass[twocolumn,floatfix, showpacs]{revtex4}

\usepackage[final]{epsfig}
\usepackage{amsmath}
\usepackage{parskip}
 
\begin{document}
 
\title{Expectations for dihadron correlation measurements at the LHC}
 
\author{Thorsten Renk}
\email{trenk@phys.jyu.fi}
\author{Kari J.~Eskola}
\email{kari.eskola@phys.jyu.fi}
\affiliation{Department of Physics, P.O. Box 35 FI-40014 University of Jyv\"askyl\"a, Finland}
\affiliation{Helsinki Institute of Physics, P.O. Box 64 FI-00014, University of Helsinki, Finland}
 
\preprint{HIP-2006-46/TH}

\begin{abstract}
The suppression of high transverse momentum ($P_T$) inclusive hadron spectra in heavy-ion collisions as compared to the scaled expectation of high $P_T$ hadron production in p-p collisions is the most direct manifestation of the interaction of hard partons with the soft bulk medium produced in heavy-ion collisions which is absent in p-p collisions. Yet the measured nuclear suppression factor $R_{AA}$ is a very averaged quantity and hence only a limited amount of information about the medium evolution and the nature of the interaction with the medium can be deduced from $R_{AA}$. Measurements of hard back-to-back hadron correlations in 200 AGeV Au-Au collisions at RHIC have opened a new window to study the energy loss of partons in a medium in a more differential way and for a different distribution of in-medium pathlengths than in the case of $R_{AA}$. In this work, we present an extrapolation of our results for back-to-back yields at RHIC energies to 5.5 TeV Pb-Pb collisions at the CERN LHC. We also discuss differences and similarities between the measurement at RHIC.

\end{abstract}
 
\maketitle

\section{Introduction}

The suppression of single inclusive high $P_T$ hadrons in heavy-ion collisions as compared to the scaled expectation from p-p collision has long been regarded as caused by energy loss of high $p_T$ partons into a dense partonic environment \cite{Jet1,Jet2,Jet3,Jet4,Jet5,Jet6}. At RHIC, the nuclear suppression factor $R_{AA}$ for pions in central Au-Au collisions has been measured out to 20 GeV \cite{PHENIX_R_AA} and a factor $\sim 5$ suppression observed.

The hope of using hard probes such as the single hadron suppression in the context of heavy-ion collisions is to infer properties of the medium and its density evolution from induced changes relative to the corresponding known hard process in p-p collisions. There is, however, growing evidence that single hadron suppression is not enough to unambiguously determine even the distribution of energy/momentum shifts of partons traversing the medium \cite{gamma-hadron, Inversion}. However, if energy loss cannot be determined reliably in momentum space, there is little hope to try to infer the QCD matter density distribution in position space.

Back-to-back correlations of hard hadrons \cite{Dijets1,Dijets2} are a more differential hard process. Due to the different geometrical averaging in the case of single hadron suppression vs. back-to-back suppression, one may hope to obtain information about the spatial distribution of dense medium from combining the two observables. While theoretical studies for back-to-back correlations as measured at RHIC have been done \cite{Correlations0,Correlations,XN-Correlations}, they seem to indicate that for RHIC kinematics the amount of additional information is not substantial, in essence various models for the energy loss mechanism and for the density evolution which describe single hadron suppression also perform well when compared with dihadron suppression. The notable exception is a class of energy loss models based on a purely linear dependence of the energy loss with pathlength --- those seem to be strongly disfavoured by back-to-back correaltions \cite{Elastic}. As suggested in \cite{Correlations}, the reason why there is only little sensitivity to the QCD matter density distribution at RHIC kinematics may be that the lever-arm in momentum is not large enough to probe substantial shifts in parton momentum --- for a steeply falling parton spectrum, even a moderate shift in parton momentum effectively resembles an absorption of partons, and this fact greatly reduces the sensitivity. At the LHC however where  the partonic $p_T$ range is large this ceases to be a problem and consequently the suppression of hard back-to-back correlated hadrons becomes a promising probe.

In this paper, we aim to provide a baseline prediction for the per-trigger yield in hard back-to-back correlations. This complements a baseline prediction of the nuclear suppression factor $R_{AA}$ \cite{R_AA_LHC} made within the same model framework. We compare with the calculation at RHIC kinematics and point out similarities and differences.

\section{The framework}

As in \cite{Correlations0,Correlations} we calculate the correlation strength of hadrons back to back with a hard trigger in a Monte-Carlo (MC) simulation. There are three important building blocks to this computation: 1) the primary hard parton production, 2) the propagation of the partons through the medium and 3) the hadronization of the primary partons. Only the step 2) probes medium properties, and hence it is here that we must specify details for the evolution of the QCD medium and for the parton-medium interaction. Let us first discuss steps 1) and 3) which are common to the simulation in p-p and Pb-Pb collisions.

\subsection{Primary parton production}

In Ref.~\cite{LOpQCD} it has been demonstrated that leading order (LO) perturbative Quantum Chromodynamics (pQCD) is rather successful in 
describing the $P_T$-spectrum of inclusive hadron production over a wide range in $\sqrt{s}$ when 
supplemented with a $\sqrt{s}$-dependent $K$-factor to adjust the overall normalization. This factor 
parametrizes next-to-leading order effects. Since we are in the following only interested in ratios 
of $P_T$-distributions, i.e. yields per triggered hadron, any factor independent of $P_T$ drops out. Hence,  in 
the following we use LO pQCD expressions without trying to adjust the absolute normalization.

The production of two hard partons $k,l$ with transverse momentum $p_T$ in LO pQCD is described by
 
\begin{equation}
\label{E-2Parton}
\frac{d\sigma^{AB\rightarrow kl +X}}{d p_T^2 dy_1 dy_2} \negthickspace = \sum_{ij} x_1 f_{i/A} 
(x_1, Q^2) x_2 f_{j/B} (x_2,Q^2) \frac{d\hat{\sigma}^{ij\rightarrow kl}}{d\hat{t}}
\end{equation}
 
where $A$ and $B$ stand for the colliding objects (protons or nuclei) and $y_{1(2)}$ is the 
rapidity of parton $k(l)$. The distribution function of a parton type $i$ in $A$ at a momentum 
fraction $x_1$ and a factorization scale $Q \sim p_T$ is $f_{i/A}(x_1, Q^2)$. The distribution 
functions are different for the free protons \cite{CTEQ1,CTEQ2} and nucleons in nuclei 
\cite{NPDF,EKS98}. The fractional momenta of the colliding partons $i$, $j$ are given by
$ x_{1,2} = \frac{p_T}{\sqrt{s}} \left(\exp[\pm y_1] + \exp[\pm y_2] \right)$.

Expressions for the pQCD subprocesses $\frac{d\hat{\sigma}^{ij\rightarrow kl}}{d\hat{t}}(\hat{s}, 
\hat{t},\hat{u})$ as a function of the parton Mandelstam variables $\hat{s}, \hat{t}$ and $\hat{u}$ 
can be found e.g. in \cite{pQCD-Xsec}. By selecting pairs of $k,l$ while summing over all allowed combinations of $i,j$, i.e. 
$gg, gq, g\overline{q}, qq, q\overline{q}, \overline{q}\overline{q}$ where $q$ stands for any of the quark flavours $u,d,s$
we find the relative strength of different combinations of outgoing partons as a function of $p_T$.

For the present investigation, we require $y_1 = y_2 = 0$, i.e. we consider only back-to-back correlations detected at midrapidity. In the first step, we sample Eq.~(\ref{E-2Parton}) summed over all $k,l$ to generate $p_T$ for the event. In the second step we perform a MC sampling of the decomposition of Eq.~(\ref{E-2Parton}) according to all possible combinations of outgoing partons $k,l$ at the $p_T$ obtained in the first step. We thus end with a back-to-back parton pair with known parton types and flavours at transverse momentum $p_T$.

To account for various effects, including higher order pQCD radiation, transverse motion of partons in the nucleon (nuclear) wave function and effectively also the fact that hadronization is not a collinear process, we fold into the distribution $\frac{d\sigma^{AB\rightarrow kl +X}}{d p_T^2 dy_1 dy_2}$ an intrinsic transverse momentum $k_T$ with a Gaussian distribution, thus creating a momentum imbalance between the two partons as ${ {p_T}_1} + { {p_T}_2} = { k_T}$. We use a value of $k_T = 2.7$ GeV for RHIC kinematics \cite{Intrinsic_k_T} and observe a sensitivity of less than 10\% of the away side yield to changes of this number by $\pm 50$\%. While $k_T$ is expected to rise at LHC conditions, in the momentum region we wish to explore, i.e. $p_T > 50 $ GeV there is no sensitivity of the calculation to the choice of intrinsic transverse momentum.

\subsection{Hadronization}

Before we can count hadrons above certain trigger or associate-momentum threshold, we have to convert the simulated partons into hadrons. More precisely, in order to determine whether there is a trigger hadron $h$ above a given threshold, given a parton $k$ with momentum $p_T$, we need to sample $A_1^{k\rightarrow h}(z_1, p_T)$, i.e. the probability distribution to find a hadron $h$ from the parton $k$ where $h$ is \emph{the most energetic hadron of the shower} and carries the momentum $P_T = z_1 \cdot p_T$.

In previous works \cite{Correlations0,Correlations} we have approximated this by the normalized fragmentation function $D_{k\rightarrow h}(z, \mu)$, sampled with a lower cutoff $z_{min}$ which is adjusted to the reference d-Au data. This procedure can be justified by noting that only one hadron with $z> 0.5$ can be produced in a shower, thus above $z=0.5$ the distributions $D_{k\rightarrow h}(z, P_T)$ and $A_1^{k\rightarrow h}(z_1, p_T)$ are  (up to the scale evolution) identical, and only in the region of low $z$ where the fragmentation function describes the production of multiple hadrons do they differ significantly. However, for a fixed cut in the hadronic momentum $P_T$, low $z$ implies probing the high $p_T$ part of a steeply falling parton spectrum, thus this region is suppressed. Empirically, sampling $D_{k\rightarrow h}(z, P_T)$ has provided a good description of the baseline \cite{Correlations}.

However, at the LHC energies, keeping the same lower cutoff $z_{min}$ is not acceptable as it would imply that there is no perturbative production of hadrons in the range below, say, 8 GeV from a 100 GeV parton. Presumably, a viable solution would be to readjust the cutoff to a measured set of reference data from p-p collisions. In the absence of such a data set, we have chosen a different path by extracting $A_1^{k\rightarrow h}(z_1, p_T)$ from shower simulations of quarks and gluons in PYTHIA \cite{PYTHIA}.

This moreover has an additional advantage, since there is a subtle point with the scale dependence of the hadronization functions. While $D_{k\rightarrow h}(z, P_T)$ is an empirical function where the scale is set by the produced hadron and the function is folded with the full parton spectrum (so that $P_T = z p_T$),  $A_1^{k\rightarrow h}(z_1, p_T)$ is an object where the scale is set by the given parton energy and the outcome of the folding is a distribution of hadron momenta. The latter is more appropriate for the situation realized in the MC simulation in which the partonic $p_T$ is fixed before hadronization.

\begin{figure}[htb]
\epsfig{file=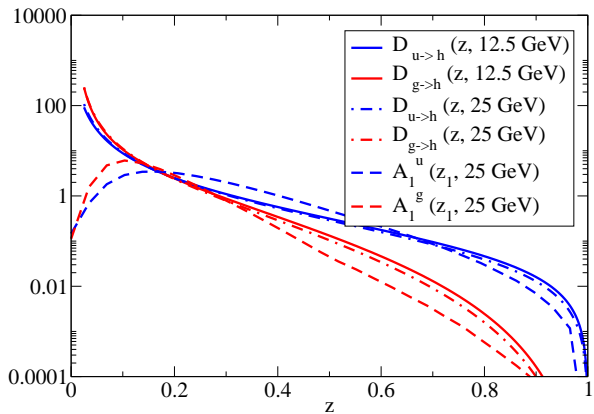, width=7.8cm}
\caption{\label{F-FFcomp}(Color online) Comparison of the KKP \cite{KKP} fragmentation function $D(z,\mu)$ for u-quarks and gluons into charged hadrons $(h^+ + h^-)$ at two hadronic scales with the leading hadron momentum fraction probability density $A_1(z_1, p_T)$ at partonic $p_T = 25$ GeV as extracted from shower simulations in PYTHIA \cite{PYTHIA}.}
\end{figure}

In Fig.~\ref{F-FFcomp} we show the extracted $A_1^{k\rightarrow h}(z_1, p_T)$ compared with $D_{k\rightarrow h}(z, \mu)$ from the KKP set of fragmentation functions \cite{KKP} in a probabilistic interpretation where we present the fragmentation function both at $\mu = 0.5 p_T$ and $\mu = p_T$, allowing a direct comparison at $z_1=0.5$ and $z_1=1$.

In the region $z < 0.5$, the quantities $A_1^{k\rightarrow h}(z_1, p_T)$ and $D_{k\rightarrow h}(z, \mu)$ in Fig.~\ref{F-FFcomp} differ from each other as expected. From the figure, it is also evident that in the region $z > 0.5$ where we expect agreement there remain differences between the empirical fragmentation function and the shower development in PYTHIA which are beyond the uncertainties introduced by the QCD scale evolution. Especially in the case of the gluon fragmentation the differences are sizeable at higher $z$. We do not wish to enter a discussion of the physics origin of these differences at this point but rather accept $A_1(z_1, \mu)$ from PYTHIA as a trial ansatz which needs to be tested against data.

Using $A_1^{k\rightarrow h}(z_1, p_T)$ on both outgoing partons from the primary pQCD vertex (see Eq.~\ref{E-2Parton}) we find the two most energetic hadrons from both showers. Following the experimental analysis procedure, out of the two we define the harder one to be the trigger hadron if it passes a set momentum cut, otherwise we ignore the event. The direction of the trigger hadron is taken to be the $(-x)$ direction and the hemisphere corresponding to this direction is referred to as 'near side', the opposite hemisphere is then defined to be the 'away side'.

In order to compute the strength of hadrons correlated with a trigger on the away side, the leading contribution is found by considering the leading hadron produced from the away side parton for which $A_1(z_1,\mu)$ is sufficient. However, in order to determine the correlation strength on the near side, the production of (at least) one more energetic hadron in the shower needs to be considered as the leading contribution here corresponds to the trigger hadron itself. Thus, note that to the same level of approximation of the shower development the near side correlation is only found at next-to-leading (NL) fragmentation whereas an away side correlation appears already at leading fragmentation and receives a correction at next-to-leading fragmentation.

In our framework, we treat this next-to-leading contribution through the conditional probability density $A_2(z_1, z_2, \mu)$ which for fixed $z_1$ corresponds to a probability density, i.e. $\int d z_1 d z_2 A_1(z_1, \mu) A_2(z_1, z_2, \mu) = 1$. This is conceptually similar to the dihadron fragmentation function introduced in \cite{DihadronFrag}. The whole hadron production in the shower arises in this language as a tower of conditional probability denities $A_N(z_1, \dots, z_n, \mu)$ with the probability to produce $n$ hadrons with momentum fractions $z_1, \dots z_n$ from a parton with energy $\mu$ being $\Pi_{i=1}^n A_i(z_1,\dots z_i,\mu)$.

In previous calculations \cite{Correlations0,Correlations} we have modelled the next-to-leading
conditional fragmentation probability using the experimentally measured probability distribution  
$A_i(z_T)$ of associated hadron production in d-Au collisions \cite{Dijets1, Dijets2} as a function  
of $z_T$ where $z_T$ is the fraction of the trigger hadron momentum carried by the associated  
hadron. A factor $\theta(E_i - E_{\rm trigger} - \Delta E - E_{\rm assoc})$ on the near  
side and $\theta(E_i - E_{\rm punch} - \Delta E - E_{\rm assoc})$ on the away side was included to
enforce energy conservation.

In this work, we use $A_2(z_1,z_2,\mu)$ as extracted from PYTHIA instead. By construction, $z_2 < z_1$ is satisfied, and energy-momentum conservation (which the extracted distribution fulfills automatically) implies $z_2 < 1 - z_1$, i.e. the region where $A_2(z_1, z_2, \mu)$ is nonzero in the $(z_1, z_2)$ plane is a triangular region from the edges  $z_1 = 0, z_1 = 1$ where $z_2 = 0$ to the center $z_1 = 0.5$ where $z_2^{max} = 0.5$.

\begin{figure*}[htb]
\epsfig{file=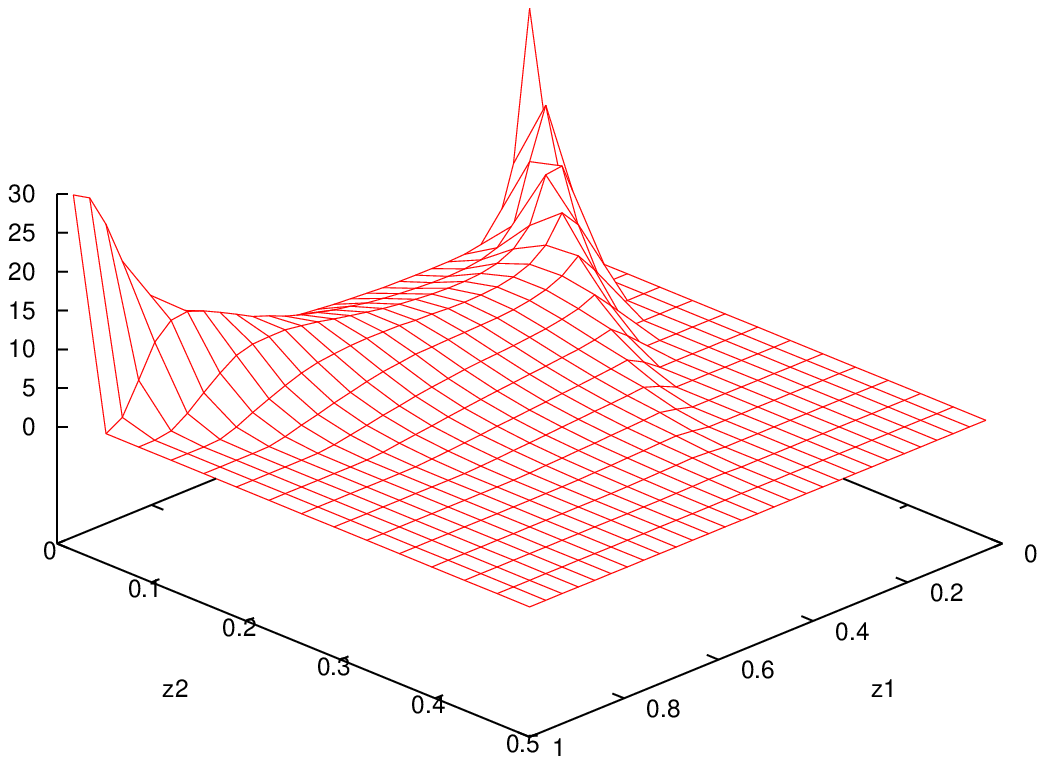, width=9.4cm} \hspace*{-1.5cm} \epsfig{file=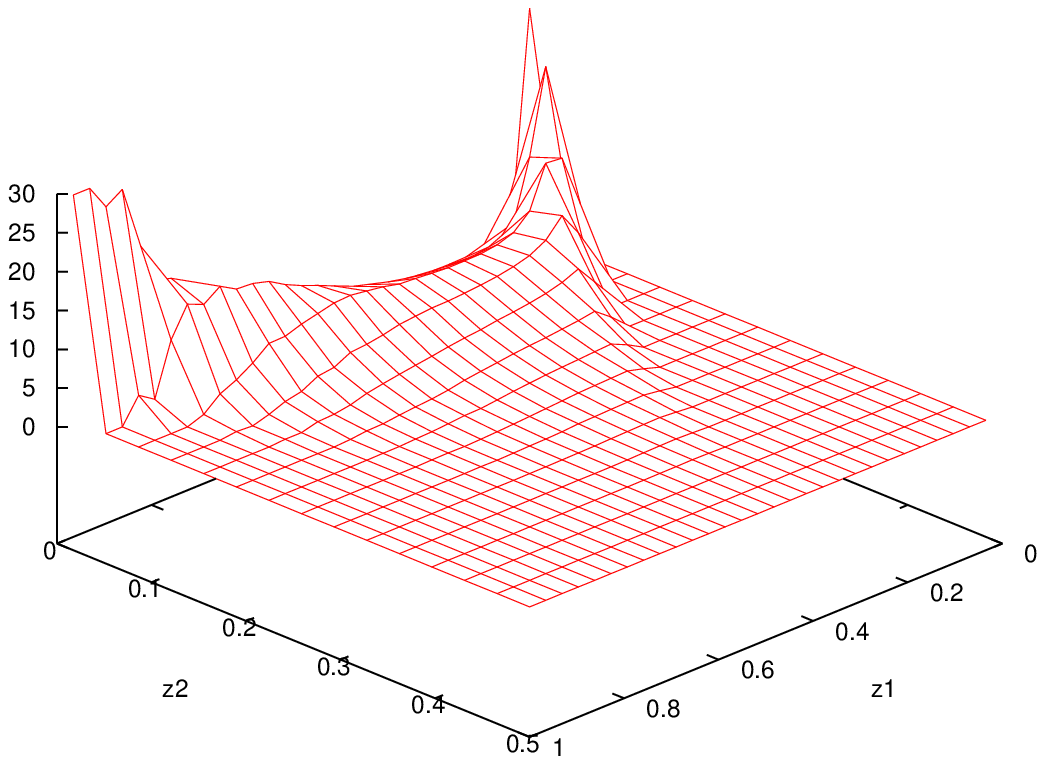, width=9.4cm} 
\caption{\label{F-A2}(Color online) Conditional probability density $A_2(z_1, z_2, \mu)$ for u-quarks (left) and gluons (right) for production of a next-to-leading hadron with momentum fraction $z_2$ given that a hadron with momentum fraction $z_1$ was produced in the shower at a parton energy scale $\mu = 25$ GeV.}
\end{figure*}

The resulting distributions for u-quarks and gluons are shown in Fig.~\ref{F-A2}.

\subsection{The medium model}

We describe the production and spacetime evolution of QCD matter in
central Pb+Pb collisions within the framework of pQCD +
saturation + hydrodynamics (the EKRT model) \cite{EKRT,ERRT,EHNRR}. In
this framework, primary parton production is computed by supplementing the
collinearly factorized pQCD parton production of Eq.~(\ref{E-2Parton})
with the conjecture of saturation of produced gluons inhibiting the growth
of gluon number at transverse momenta below the saturation scale,
$p_T<p_{\rm sat}$. In central heaviest-ion collisions at RHIC and LHC, the
computation remains in the perturbative regime as $p_{\rm sat}\approx
1.2-2.0$~GeV $\gg \Lambda_{\rm QCD}$. The average formation time of the
QCD system, $\tau_0\sim 1/p_{\rm sat}=0.17-0.2$~fm/$c$ at RHIC and LHC,
and pQCD parton production in a central rapidity unit can then be
computed, and thus the initial QGP densities estimated, see
\cite{ERRT,EHNRR} for details.
 
With the assumptions of full thermalization and no initial transverse flow
at $\tau_0$, we then apply relativistic, longitudinally boost-symmetric,
azimuthally symmetric, longitudinally and transversally expanding ideal
hydrodynamics to describe the spacetime evolution of the formed QCD
matter.
We use the results obtained in \cite{EHNRR}, where the QGP is assumed to
be an ideal gas of gluons and three flavours of massless quarks and
antiquarks, while the
hadron resonance gas consists of all hadronic states up to mass 2 GeV. A
first-order phase transition takes place at a critical temperature 165
MeV, and, with binary-collision initial energy-density profiles, the
system decouples at $T_{\rm dec}=150$ MeV (see also the discussion of the
decoupling dynamics and $T_{\rm dec}$ in \cite{ENR07}). Illustrations of
the phase boundaries, flow lines and transverse velocity contours for
central Au+Au collisinons at RHIC and Pb+Pb at LHC, as well as extensive
comparison with the hadron spectra measured at RHIC and further discussion
can be found in \cite{EHNRR}.
 
Within the framework we apply here, there are obvious theoretical
uncertainties in the extrapolation to the LHC energies, which are related
to the gluon shadowing (see \cite{EKPS}) in the primary production, and to
the QCD matter equation of state and to the currently debated possible
dissipative effects \cite{HEINZ, ROMATSCHKE} in the matter evolution
itself. Accepting these uncertainties, however, the benefit in using the
EKRT model is that being a closed framework it has predictive power: once
$T_{\rm dec}$ has been fixed at RHIC \cite{EHNRR}, straightforward
RHIC-tested predictions for the LHC hadron spectra can be made
\cite{EHNRR,EHNRR_CERN}, and the evolution of QCD matter densities and
flow velocities, the input for the present paper, obtained.

\subsection{Parton-medium interactions}

In the case of heavy-ion collisions, partons emerging from a hard vertex are surrounded by the soft bulk matter and may lose
energy before hadronization.
We compute the energy loss of a hard parton traversing a thermalized medium using the formalism by Baier, Dokshitzer, Muller, Peigne and Schiff (BDMPS) \cite{Jet2} which assumes that energy is predominantly lost due to medium-induced radiation. This formalism is cast into the form of
energy loss probability distributions, so-called quenching weights in \cite{QuenchingWeights}. This is a convenient formulation for the
purpose of a MC simulation.

The key quantity characterizing the energy loss induced by a medium with energy density $\epsilon$ 
in the BDMPS formalism \cite{Jet2} is the local transport coefficient $\hat{q}(\tau, \eta_s, r)$ which 
characterizes the squared average momentum transfer from the medium to the hard parton per unit 
pathlength. Since we consider a time-dependent inhomogeneous medium, this quantity depends on 
proper time $\tau = \sqrt{t^2-z^2}$, spacetime rapidity $\eta_s = \frac{1}{2}\ln \frac{t+z}{t-z}$,
cylindrical radius $r$ and in principle also on azimuthal angle $\phi$, but for the time 
being we focus on central  collisions only.
 
The transport coefficient is related to the energy density of the medium $\epsilon$ as
 
\begin{equation}
\label{E-qhat}
\hat{q}(\tau, \eta_s, r) = K \cdot 2 \cdot [\epsilon(\tau, \eta_s, r)]^{3/4} (\cosh \rho - \sinh \rho \cos\alpha)
\end{equation}
 
with $K=1$ for an ideal quark-gluon plasma (QGP) \cite{Baier} Here, $\rho$ is the local flow rapidity with angle $\alpha$ between flow and parton trajectory \cite{Flow1,Flow2}. The parameters $\epsilon$ and  $\rho$  characterize the medium and must be supplied by
the underlying QCD medium evolution model. 

In the following, motivated by 
Ref.~\cite{Baier} we assume the proportionality constant $K$ to remain unaltered in different 
phases of the medium. We treat $K$ as an adjustable parameter (for a detailed discussion see \cite{Correlations}) which 
however, once fixed, is independent on $\sqrt{s}$ of the heavy-ion collision and assumes the same value at RHIC and LHC 
energies while the medium density given by $\epsilon(\tau, \eta_s, r)$ changes substantially.
 
Given the local transport coefficient at each spacetime point, a parton's energy loss depends on 
the position of the hard vertex at ${\bf {r}_0} = (x_0,y_0)$ in the transverse plane 
at $\tau=0$ and the 
angular orientation of its trajectory $\phi$ (i.e. its path through the medium). In order to determine the probability  $P(\Delta E,  
E)_{path}$ for a 
hard parton with energy $E$ to lose the energy $\Delta E$ while traversing the medium on its 
trajectory, we make use of a scaling law \cite{JetScaling} which allows to relate the dynamical 
scenario to a static equivalent one by calculating the following quantities
averaged over the jet trajectory $\xi(\tau):$
 
\begin{equation}
\label{E-omega}
\omega_c({\bf r_0}, \phi) = \int_0^\infty d \xi \xi \hat{q}(\xi)
\end{equation}
and
\begin{equation}
\label{E-qL}
\langle\hat{q}L\rangle ({\bf r_0}, \phi) = \int_0^\infty d \xi \hat{q}(\xi)
\end{equation} 
 
as a function of the jet production vertex ${\bf r_0}$ and its angular orientation $\phi$. We set 
$\hat{q} \equiv 0$ whenever the decoupling temperature of the medium $T = T_F$ is reached.

Using the numerical results of \cite{QuenchingWeights}, we obtain $P(\Delta E)_{path}$ 
for $\omega_c$ and $R=2\omega_c^2/\langle\hat{q}L\rangle$ 
as a function of jet production vertex and the angle $\phi$ corresponding to
 
\begin{widetext}
\begin{equation}
P(\Delta E)_{path} = \sum_{n=0}^\infty \frac{1}{n!} \left[ \prod_{i=1}^n \int d \omega_i  
\frac{dI(\omega_i)}{d \omega}\right]
\delta\left( \Delta E - \sum_{i=1}^n \omega_i\right) \exp\left[-\int d\omega\frac{dI}{d\omega}  
\right]
\end{equation}
 
\end{widetext}
 
which makes use of the distribution $\omega \frac{dI}{d\omega}$ of gluons emitted into the jet  
cone.
The explicit expression of this quantity for the case of multiple soft scattering can be found in  
\cite{QuenchingWeights}. Note that the formalism of \cite{QuenchingWeights} is defined for the  
limit of asymptotic parton energy, hence the probability distribution obtained $P(\Delta E)_{path}$  
is independent of $E$.
 
The initially produced hard parton spectrum and, consequently, the number of hard vertices
in the $(x,y)$ plane (where the $z$-axis is given by the beam direction) in an 
$A-A$ collision at fixed impact parameter {\bf b}, are proportional to the nuclear overlap, 
 
\begin{equation}
\frac{dN_{AA}^f}{dp_T^2dy_f} = T_{AA}({\bf b})\frac{d\sigma^{AA\rightarrow f+X}}{dp_T^2 dy_f},
\label{inipartons}
\end{equation}
 
where $T_{AA}({\bf b})$ is the standard nuclear overlap function. 
We define a normalized geometrical distribution $P(x_0,y_0)$ for central collisions as

\begin{equation}
\label{E-Geo}
P(x_0,y_0) = \frac{[T_{A}({\bf r_0})]^2}{T_{AA}(0)},
\end{equation}
 
where the thickness function is given in terms  of the nuclear density 
$\rho_A({\bf r},z)$ as $T_A({\bf r})=\int dz \rho_A({\bf r},z)$. 

For a given event in the MC simulation, we first sample Eq.~(\ref{E-Geo}) to determine the vertex position in the transverse plane. From this point, we propagate the parton through the medium and we evaluate Eqs.~(\ref{E-omega},\ref{E-qL}) along the path, on which the medium properties enter via Eq.~(\ref{E-qhat}). Computing the corresponding $P(\Delta E)_{path}$ for each parton, we sample this distribution to determine the actual energy loss for near and away side parton in the event. In the case $\Delta E > E_{parton}$ we consider the parton to be absorbed by the medium and not contributing to high $p_T$ yield.

\section{Results}

\subsection{Comparison with 200 AGeV d-Au data}

\begin{figure*}[htb]
\epsfig{file=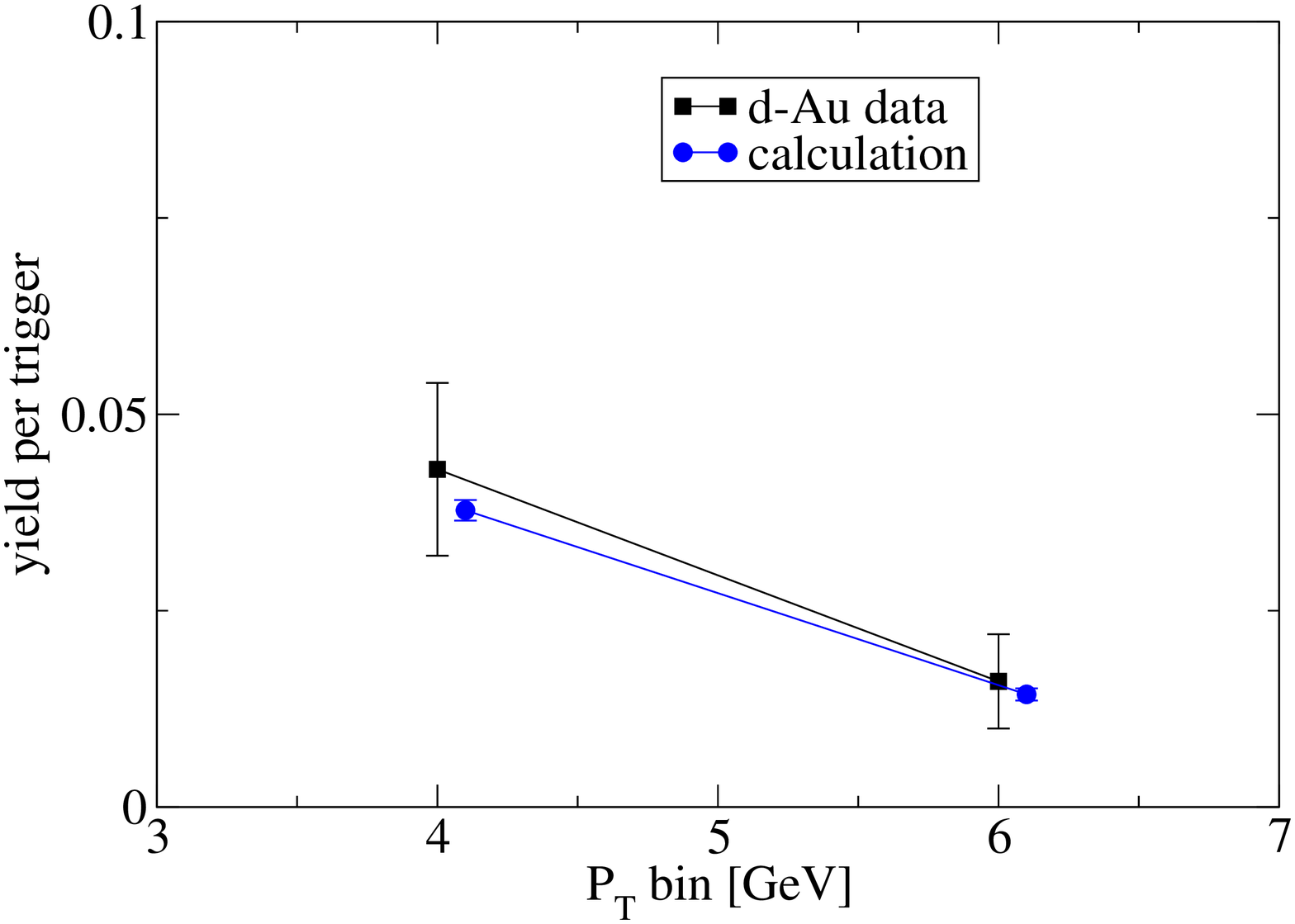, width=7.5cm}\epsfig{file=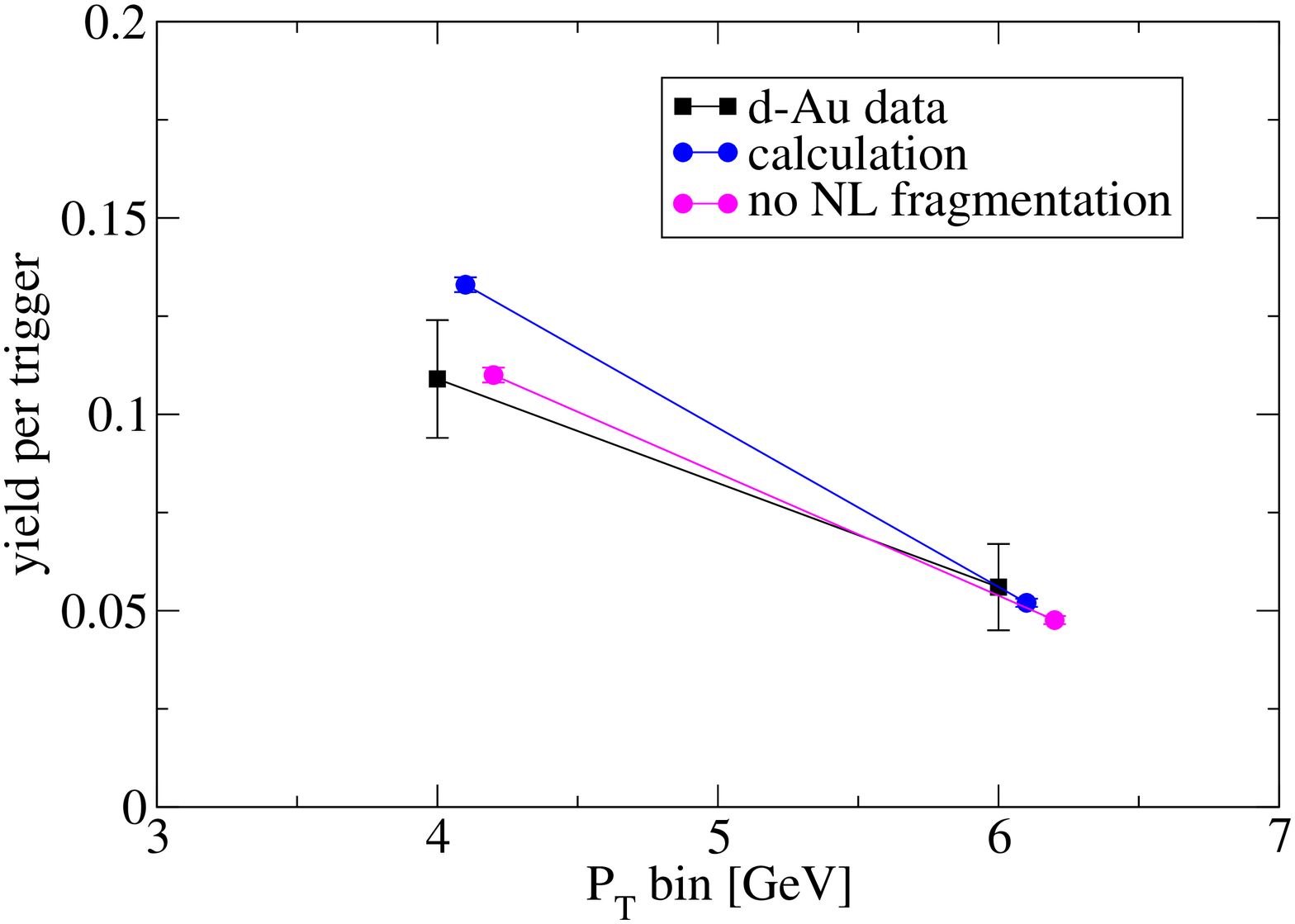, width=7.5cm}
\caption{\label{F-Baseline}(Color online) Yield per trigger-hadron between 8 and 15 GeV in d-Au collisions at 200 AGeV on near side (left panel) and away side (right panel) as obtained by the STAR collaboration \cite{Dijets1,Dijets2} and compared with the MC simulation described in
this work for two associate momentum bins of 4-6 GeV and 6+ GeV. The different data points have been shifted artificially along
the $x$-axis for clarity of presentation. The right panel also includes the result neglecting the NL fragmentation of the away-side parton.}
\end{figure*}

The first step is to test the MC simulation without a medium to see how well LO pQCD combined with the first two terms
in the hadronization probability distribution expansion, i.e. $A_1(z_1, \mu)$ and $A_2(z_1, z_2, \mu)$ describes the data.
For this purpose, we compare in Fig.~\ref{F-Baseline} with the correlation strength measured in d-Au collision by the STAR collaboration \cite{Dijets1,Dijets2}.
The data have been taken for a trigger hadron between 8 and 15 GeV and in two associate yield bins (on both near and away side) of
4-6 and 6+ GeV. In order to simulate d-Au collisions, we insert a nuclear parton distribution function into Eq.~(\ref{E-2Parton}) and 
assume that no medium is formed and that nuclear matter is not dense enough to induce significant energy loss.

As apparent from the figure, the calculation is able to describe the data generally rather well, however in the 4-6 GeV momentum bin on the away side it overshoots the measured data. We have re-run the simulation taking into account only the leading fragmentation term $A_1(z_1, \mu)$ and find that this contribution alone seems to do well with the data.

Since the trigger is always chosen to be the most energetic hadron in the simulation, there is considerable bias towards the trigger parton being a quark, as the quark fragmentation is harder (cf. Fig.~\ref{F-FFcomp}). In the kinematical range probed, the $q(\overline{q}) g \rightarrow q(\overline{q}) g$ subprocess is rather important, thus there is a bias towards a gluon being the away side parton. We note from Fig.~\ref{F-FFcomp} that the discrepancy between the gluon fragmentation function and $A_1^{g\rightarrow h}(z_1, \mu)$ is larger than for quarks. Quite likely, this discrepancy also propagates into the extraction of $A_2^{g\rightarrow h} (z_1, z_2, \mu)$ from PYTHIA and is the reason that the present calculation shows differences here relative to the data (and to previous results \cite{Correlations0,Correlations}).

The eventual solution to this discrepancy would be to find a different way to extract $A_1(z_1, \mu)$ and $A_2(z_1, z_2, \mu)$ than PYTHIA shower simulations. However, since the overall description of the data seems to be reasonable, we proceed with the hadronization procedure as introduced above, but assign an uncertainty to the simulations of the heavy-ion case which we estimate by calculating the results with and without $A_2(z_1, z_2, \mu)$ on the away side.

\subsection{Comparison with 200 AGeV central Au-Au data}

\begin{figure*}[htb]
\epsfig{file=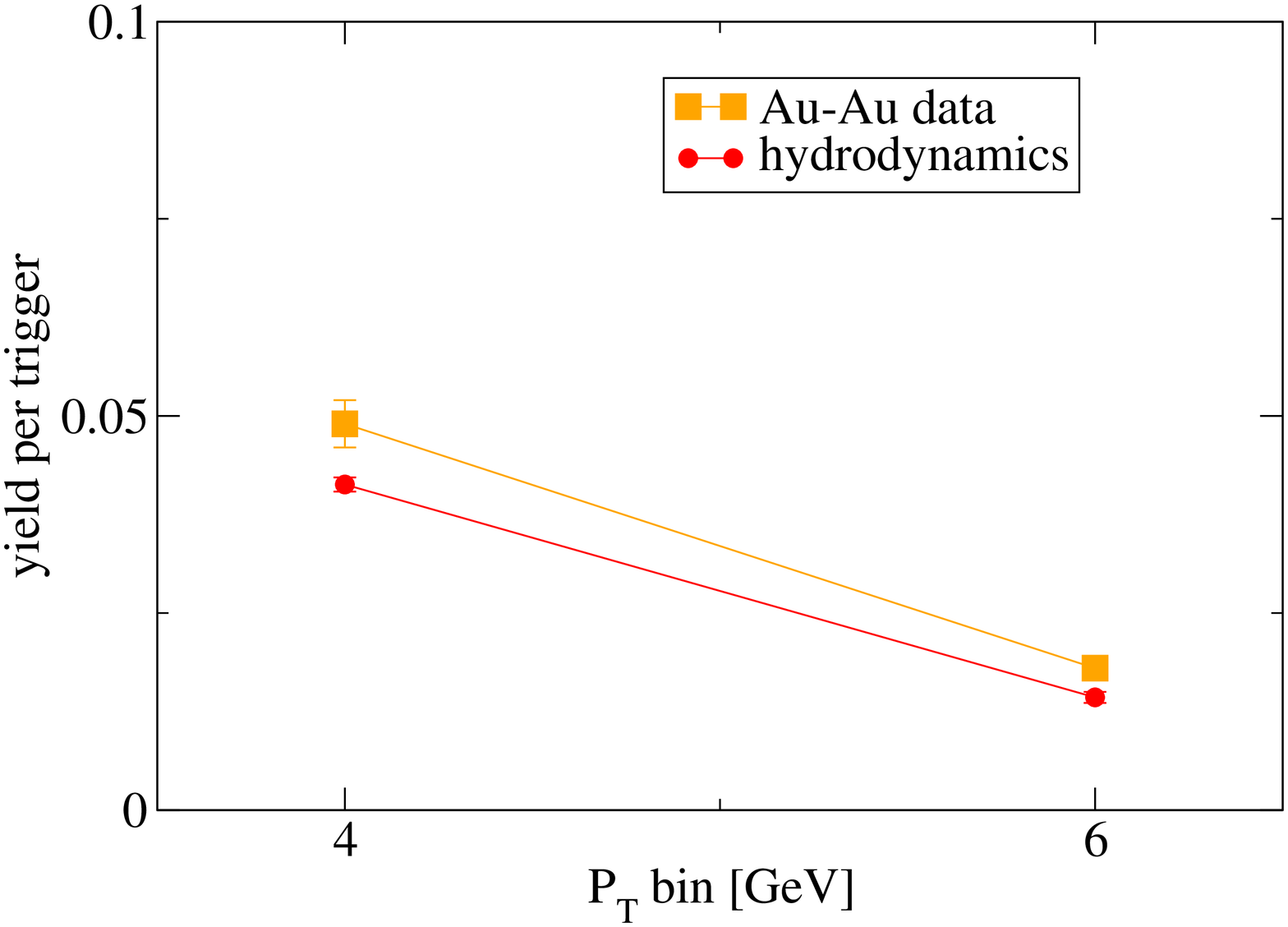, width=7.5cm} \epsfig{file=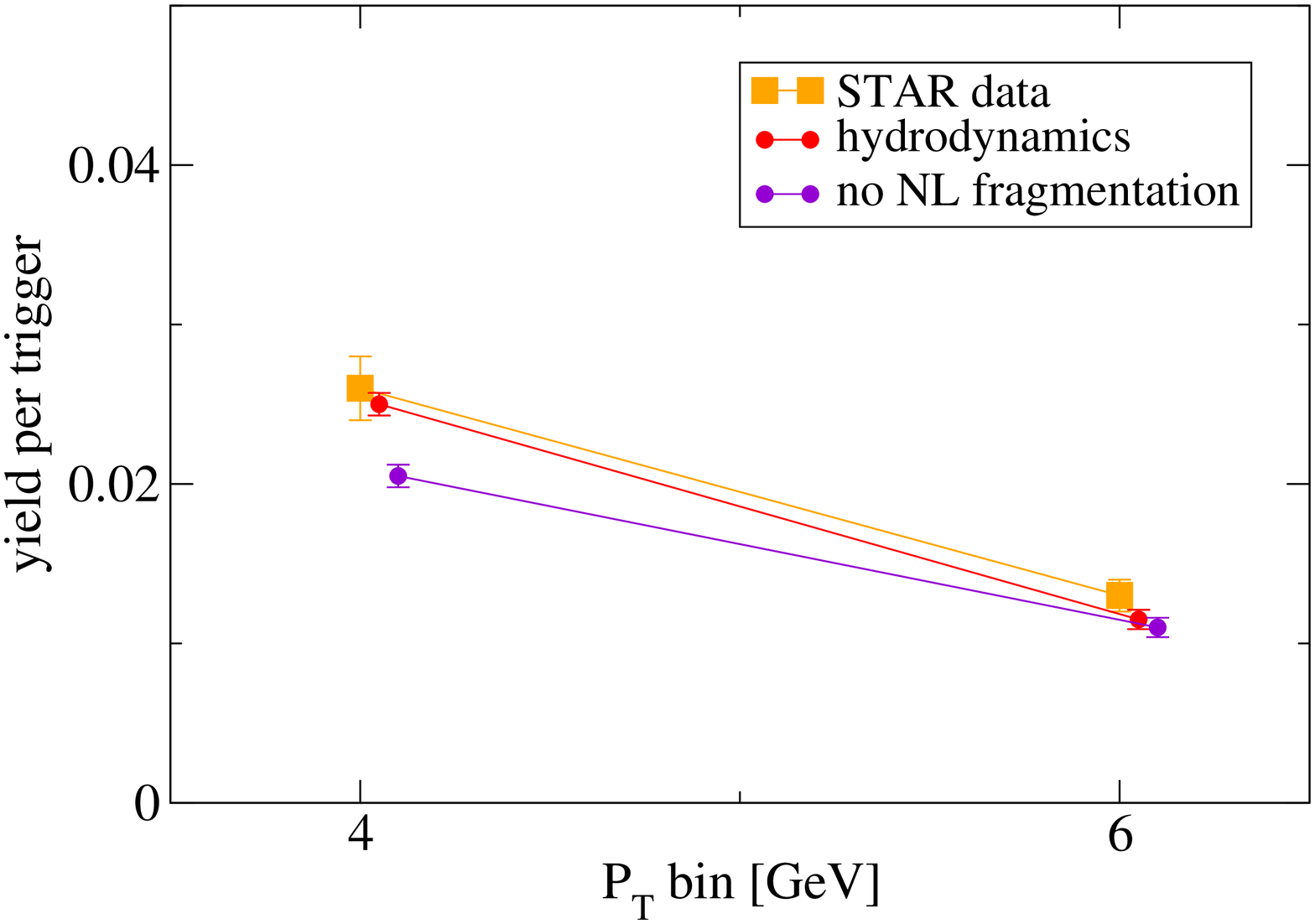, width=7.5cm}
\caption{\label{F-AuAu}(Color online) Yield per trigger hadron between 8 and 15 GeV in central Au-Au collisions at 200 AGeV on near side (left panel) and away side (right panel) as obtained by the STAR collaboration \cite{Dijets1,Dijets2} and compared with the MC simulation with a hydrodynamical evolution for the soft medium background described in
this work for two associate momentum bins of 4-6 GeV and 6+ GeV. The different data points have again been shifted artificially along
the $x$-axis for clarity. The right panel also includes the result neglecting the NL fragmentation of the away side parton.}
\end{figure*}

In the next step, we test the MC simulation including in-medium energy loss with a soft medium background evolution as provided by
a hydrodynamical calculation for 200 AGeV central Au-Au collisions. The parameter $K$ in Eq.~(\ref{E-qhat}) is fixed by the single
hadron suppression for this system \cite{Correlations}, thus the computation proceeds without any further free parameters. The resulting yields per trigger as a function of associate hadron momentum for near and away side are shown in Fig.~\ref{F-AuAu}.

We observe that the yield on the near side is almost identical to the calculated yield in d-Au collisions and systematically below the data. One way of resolving the discrepancy while retaining agreement with the d-Au data could be the inclusion of NNL fragmentation processes. Due to the large errors in the d-Au measurement, this would not destroy the agreement with the baseline. While a full computation of the NNL conditional fragmentation probability distribution $A_3(z_1, z_2, z_3, \mu)$ is very involved, we have checked that the approximation 

\begin{equation}
A_3(z_1, z_2, z_3, \mu) \approx A_2(z_1+z_2, z_3, \mu) \theta(z_2 - z_3)
\end{equation}

produces the right order of magnitude to agree with the data. A different possibility is a sizeable contribution of coalescence processes \cite{Coalescence,Coalescence2} in the Au-Au case to the near side hadron yield which is not included in the calculation. If coalescence processes are relevant, one would expect to systematically underestimate the data. Given the large uncertainty of the baseline d-Au measurement and the theoretical uncertainty in the shower modelling using PYTHIA, it is unfortunately impossible to make a definite statement what the relevant physics process underlying the discrepancy is.

On the away side, the calculation shows good agreement in both the 4-6 GeV and the 6+ GeV momentum bin. However, given that the calculation overshoots the d-Au baseline in the lower momentum bin, one has to conclude that the agreement in this bin is accidental.

This is clearly confirmed by omitting NL fragmentation on the away side (which restores agreement with the baseline): doing so 
results in an underprediction of this bin just as observed in our previous calculation \cite{Correlations} which is consistent
with the observation that coalescence processes \cite{Coalescence,Coalescence2,Reco} play a substantial role for hadronization in this momentum region.

\begin{figure}[htb]
\epsfig{file=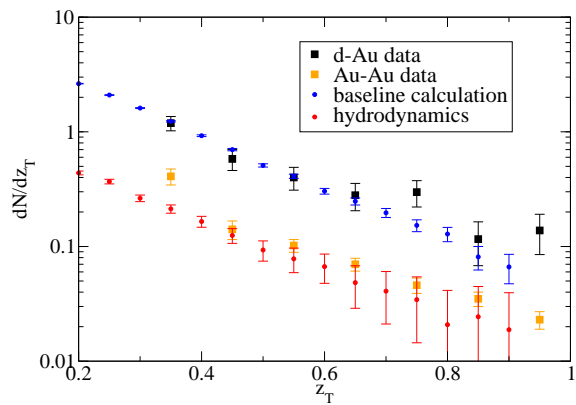, width=7.5cm}
\caption{\label{F-z_T}(Color online) $z_T$ distribution of away side associated hadrons measured in 200 AGeV d-Au and central Au-Au collisions in comparison with the MC simulation results. Here, 'baseline calculation' (see text) denotes the simulation without a medium and 'hydrodynamics' the simulation carried out using a hydrodynamical evolution model for the QCD matter produced.} 
\end{figure}

In Fig.~\ref{F-z_T}, we show the distribution of $z_T = P_T^{assoc}/P_T^{trigger}$ on the away side in d-Au and central Au-Au collisions and compare with data. We observe a general tendency that the calculation slightly underestimates the data in the high $z_T$ region. This is consistent with the observation that there is a trend that the calculation may underestimate the higher associate momentum window
slightly. However, within errors, the results are compatible.

\subsection{Prediction for 5.5 ATeV central Pb-Pb collisions}

We compute the expected correlation strength pattern for an energy of 5.5 ATeV for p-p collisions and Pb-Pb collisions as expected to take
place at the CERN LHC. For the trigger range, we assume 50-70 GeV to probe a regime which can be accessed at LHC but is beyond the reach of the RHIC experiment.

Since our formalism includes shower development only to the second term $A_2(z_1, z_2, \mu)$, the associate hadron momentum cuts cannot be too low, otherwise the main contribution to hadron production cannot be assumed to be dominated by the leading and next to leading hadron. Thus, we start considering associate hadron production for 20 GeV and above up to the trigger energy in 5 GeV momentum windows. In order to account for the uncertainty with regard to the NL fragmentation on the away side as discussed earlier, we carry out the simulation both including and excluding this contribution to provide an error estimate.

\begin{figure*}
\epsfig{file=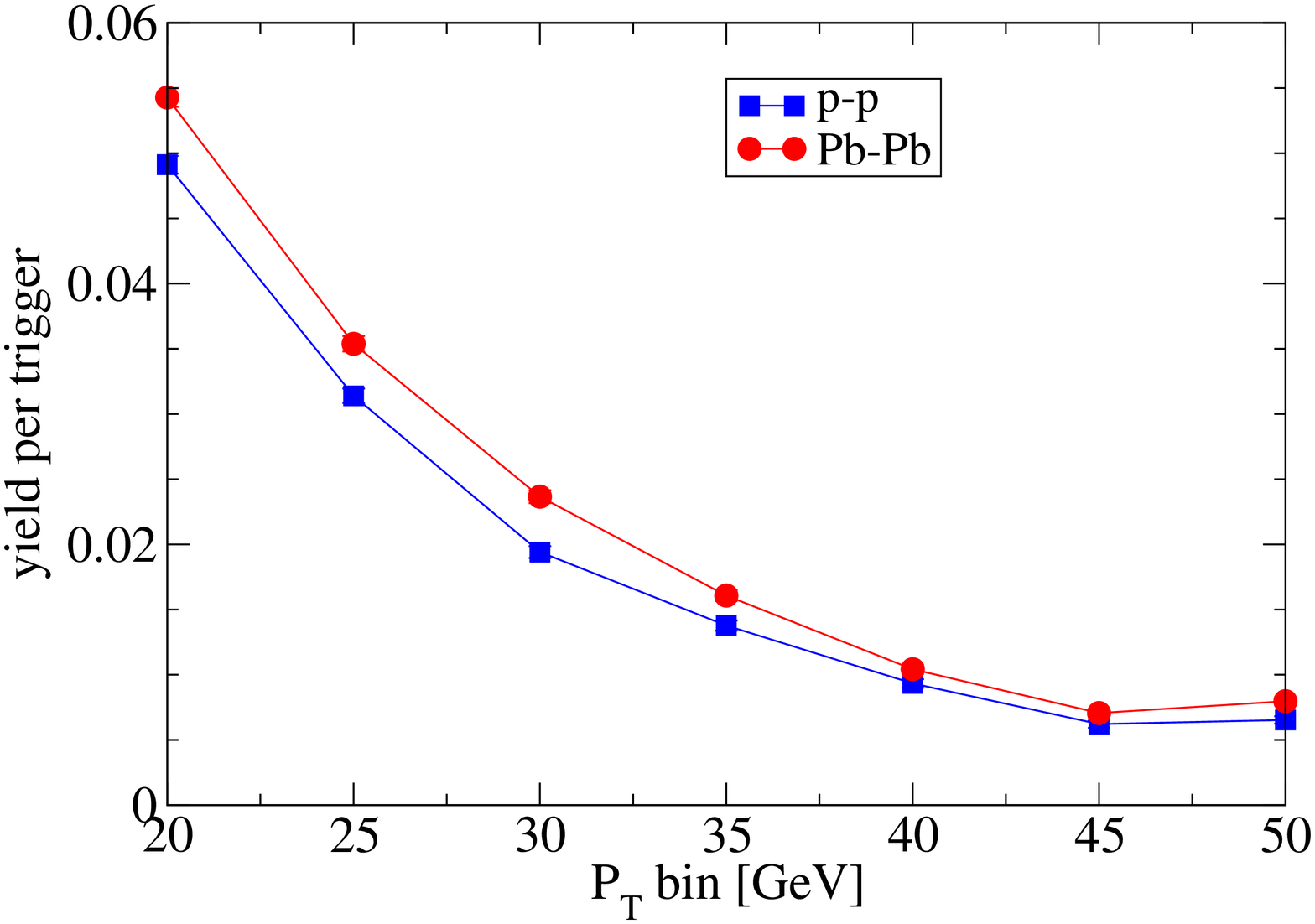, width=7.5cm} \epsfig{file=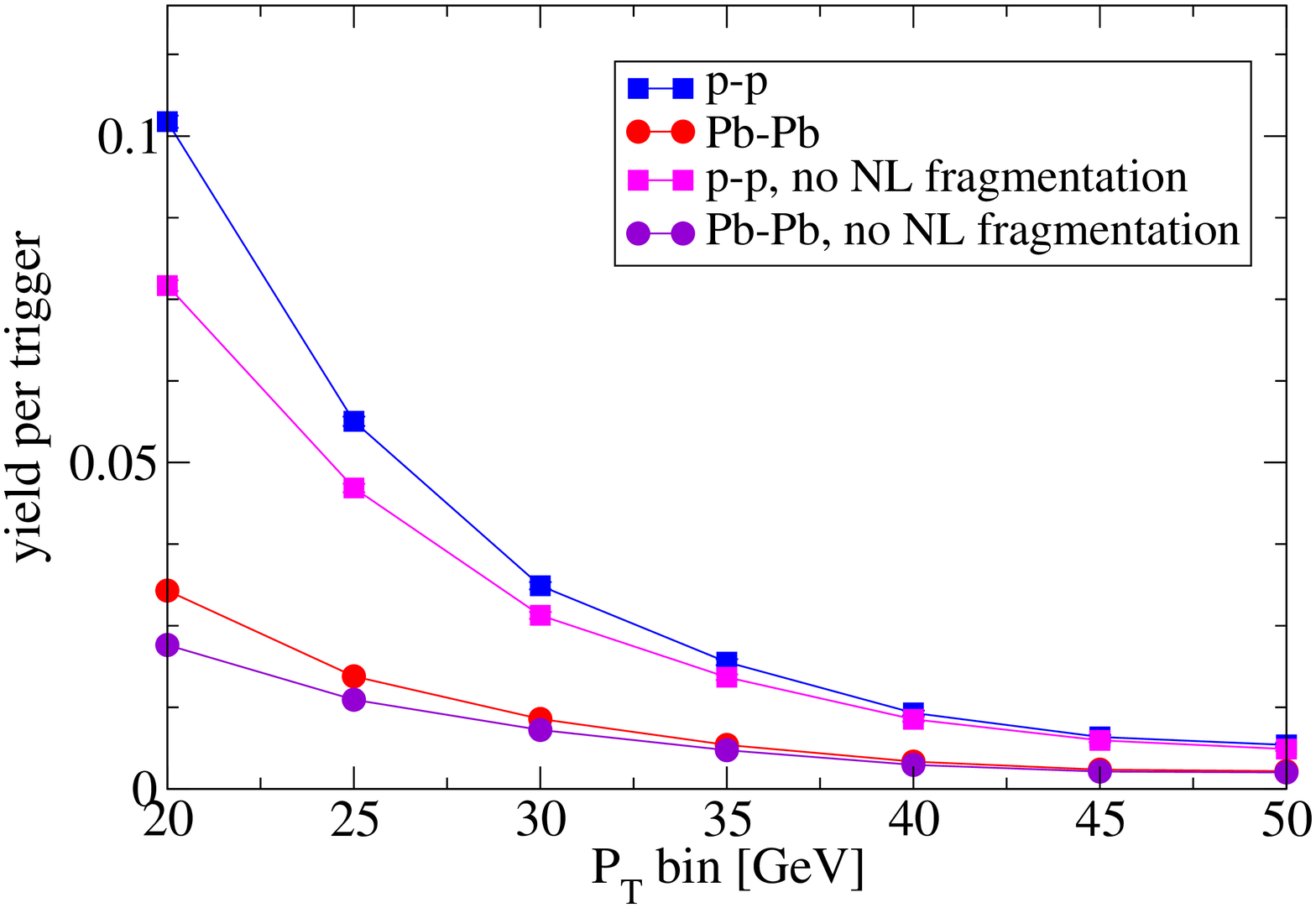, width=7.5cm}
\caption{\label{F-PbPb}Yield per trigger hadron between 50 and 70 GeV in p-p and central Pb-Pb collisions at 5.5 ATeV on near side (left panel) and away side (right panel) as obtained in our MC simulation with a hydrodynamical QCD medium evolution as a function of associate momentum bin. The right panel also includes the result neglecting the NL fragmentation of the away-side parton.}
\end{figure*}

In Fig.~\ref{F-PbPb} we show the resulting yield per trigger on the near and away side. Let us first focus on the away side (right panel). Regardless if one uses NL fragmentation contributions or not, there is a suppression of the away-side yield when going from p-p to Pb-Pb collisions of about a factor 3. This is comparable with the factor 4 suppression observed at RHIC. The similarity is not quite unexpected, since in the range of the trigger momentum the same model predicts a single hadron suppression factor of about the magnitude seen at RHIC \cite{R_AA_LHC}. The underlying reason is that the ratio between the typical scale of energy loss vs. the momentum scale probed is similar and the change in characteristic length scales between central Au-Au and central Pb-Pb is negligible. Up to a momentum scale of about 35 GeV, there is an uncertainty associated with NL fragmentation, for higher momenta the uncertainty goes away .

Let us now focus on the near side (left panel in Fig.~\ref{F-PbPb}): Here we observe that the yield per trigger is enhanced by 10-20\% in Pb-Pb relative to to p-p collisions. At first glance, this seems to be surprising. However, one should keep in mind that the model still includes a strong suppression of trigger hadrons, only the per-trigger yield is enhanced. These findings can be reconciled by the observation that the dense medium at LHC acts like a gluon filter and in essence enhances the relative abundance of quarks escaping the medium to gluons escaping the medium. The underlying reason is that gluons, due to their different color charge, couple more strongly to the medium than quarks and consequently lose more energy. However, due to the harder leading and NL fragmentation, quark jets on average produce harder associated yield than gluon jets, which is precisely what is seen in the simulation.

\begin{figure}
\epsfig{file=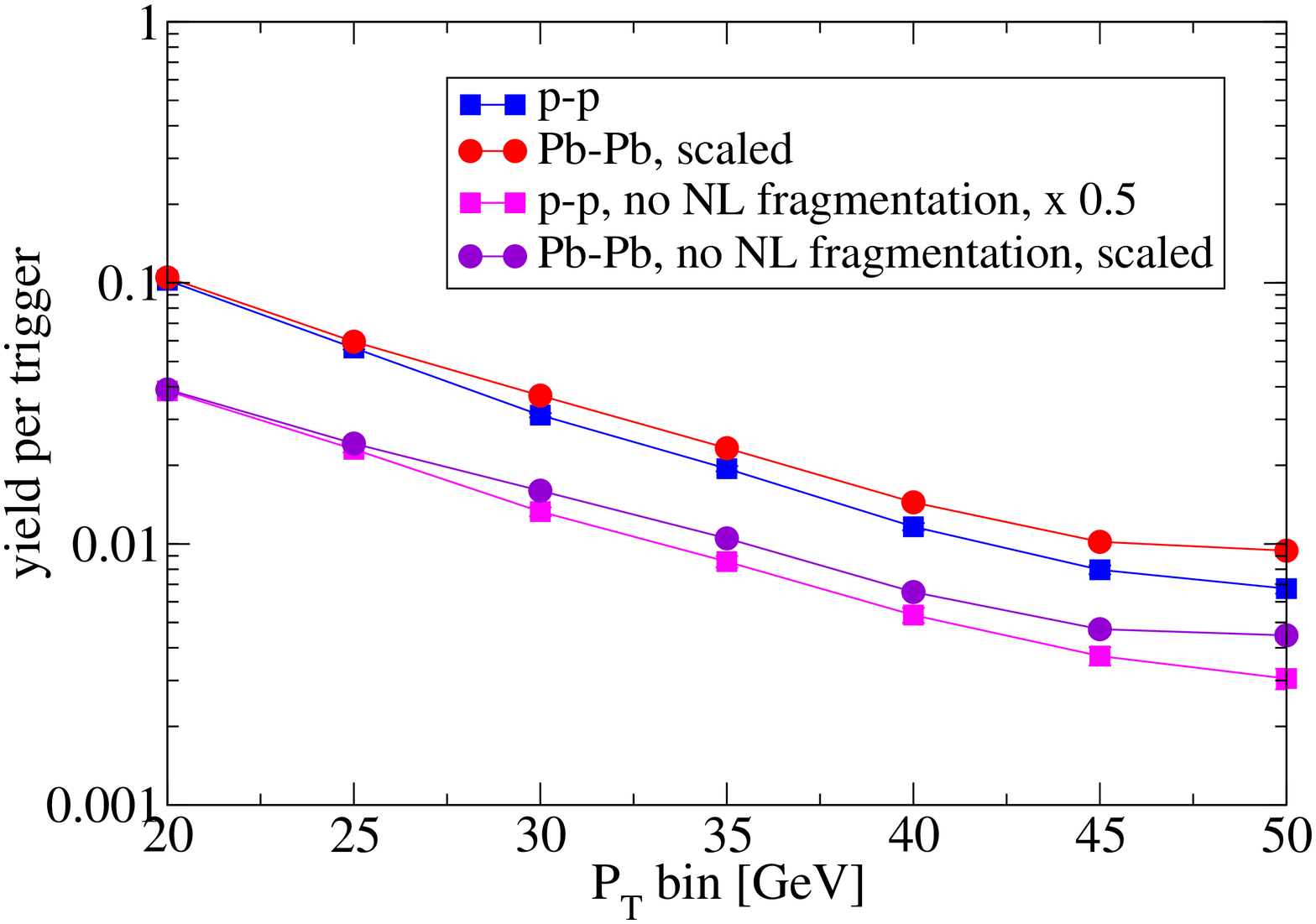, width=7.5cm}
\caption{\label{F-pshift}(Color online) Yield per trigger hadron between 50 and 70 GeV in {p-p} and central Pb-Pb collisions at 5.5 ATeV on the away side as obtained in our MC simulation. The yield in both Pb-Pb cases is scaled with a constant to agree with the first point of the p-p computation. The distortion of the momentum spectrum of recoil hadrons due to energy loss is clearly visible.} 
\end{figure}

Finally, in Fig.~\ref{F-pshift} we take a closer look at the momentum spectrum of hadrons recoiling from the trigger. To first order, one finds the reduction by a factor three mentioned above when going from p-p to Pb-Pb collisions. This would be consistent with complete absorption of 2/3 of all partons whereas 1/3 passes the medium without any energy loss. However, as a closer comparison reveals, there is a momentum dependent distortion of the spectra in heavy-ion collisions which is associated with events in which an away-side parton emerges from the medium after a finite energy loss $\Delta E$. From this $p_T$ dependent shift of the spectrum, one can in principle derive constraints about the averaged energy loss probability $P(\Delta E, E)$ and hence about the medium density distribution.

\section{Discussion}

We have made a baseline prediction for the expected per-trigger yield of associate hadrons as a function of transverse momentum in central Pb-Pb collisions at 5.5 ATeV for a trigger energy of 50 - 70 GeV. We have demonstrated that the effect of the medium is expected to leave a characteristic imprint in the observables. 

On the near side, we expect enhancement of the associate hadron yield as compared to the p-p case due to the fact that energy loss of gluons is stronger, hence there is some bias towards quarks leading to a trigger hadron, and the fact that the NL fragmentation for quarks is harder leads to a slight enhancement in the relevant momentum region.

On the away side, we expect a substantial suppression of the yield as compared to the p-p case, although the suppression we find is in relative terms smaller than what is currently observed at RHIC. We have further shown that the shape of the momentum spectrum of away-side hadrons is not equal to the momentum spectrum expected for p-p collisions. This is because the medium does not only act by absorbing partons, but one gets a chance to observe the characteristic shift in parton energy induced by the medium. This allows a more detailed view on the geometry-averaged energy loss probability distribution. In this respect, the observable shows a potential which may not be accessible at RHIC kinematics.

There are multiple uncertainties associated with this prediction at each step of the MC simulation. In the LO pQCD calculation, both NLO corrections and the choice of the nuclear parton distribution functions result in an uncertainty, especially as the role of gluon shadowing cannot be reliably extrapolated to the LHC regime. However, we have tested that the difference between the NPDF \cite{NPDF} and the EKS98 \cite{EKS98} nuclear parton distribution sets is smaller than the statistical errors of the MC simulation. While the yield of partons at given $p_T$ is sensitive to the nuclear parton distribution, the associated per-trigger yield is not since many uncertainties cancel in the ratio. 

There is likewise an uncertainty associated with the choice of the hadronization scheme. We have tried to show the magnitude of this uncertainty by comparing with available RHIC data and by indicating the relative magnitude of the leading and next-to-leading fragmentation term in the prediction.

A large uncertainty must also be assigned to the hydrodynamical prediction of the underlying medium evolution. At present, neither entropy production and the distribution in the initial state nor the equilibration process or the equation of state governing a hydrodynamical expansion are precisely known. Yet all these quantities would leave a characteristic signature on the geometrical bias induced by the requirement to obtain a trigger. 

Keeping in mind these uncertainties,  the prediction presented here should in any case show the right order of magnitude of the correlation strength. Furthermore, we believe the essential effects are described qualitatively correct, given that the radiative energy loss picture is valid.

\begin{acknowledgments}
 
This work was financially supported by the Academy of Finland, Project 115262. 
 
\end{acknowledgments}

\end{document}